\documentclass[12pt]{article} 
\usepackage[english]{babel}
\usepackage{bbold}
\usepackage{soul}
\usepackage{amsmath}
\usepackage{amsthm}
\usepackage{amscd}
\usepackage{amssymb}
\usepackage{dsfont}
\usepackage{bm}
\usepackage{caption}
\usepackage{leftidx}
\usepackage[makeroom]{cancel}
\usepackage{indentfirst}
\usepackage{hyperref}
\usepackage[com pat=1.1.0]{tikz-feynman}
\usepackage{subfig}
\usepackage{stackengine}
\usepackage{enumitem}
\usepackage{courier}
\usepackage{graphicx}
\usepackage{mathtools}
\usepackage{euscript}
\usepackage[titletoc,title]{appendix}

\usepackage{adjustbox}
\usepackage{mathtools}
\usepackage{multirow}
\usepackage{comment}
\usepackage{geometry}
\newcommand {\ket} [1] {\left| #1 \right\rangle}

\newcommand{\bma} {\begin{pmatrix}}
\newcommand{\ema} {\end{pmatrix}}

\newtheoremstyle{dotless}{}{}{\itshape}{}{\bfseries}{}{ }{}
\theoremstyle{dotless}

\newcommand{\slashD}{/\!\!\!\!{D}}

\newcommand{\s}{\nobreak\hspace{.11em}\nobreak}
\newcommand{\beq}{\begin{equation}}
\newcommand{\eeq}{\end{equation}}
\newcommand{\beqn}{\begin{eqnarray}}
\newcommand{\eeqn}{\end{eqnarray}}

\newcommand{\bpma} {\begin{pmatrix*}[r]}
\newcommand{\epma} {\end{pmatrix*}}

\newcommand{\tr}{\text{Tr}}
\newcommand{\cpn}{$\mathbb{CP}(N)$ }
\newcommand{\cpnm}{$\mathbb{CP}(N-1)$ }

\usepackage{authblk}
\makeatletter
\renewcommand\AB@authnote[1]{\rlap{\textsuperscript{\normalfont#1}}}

\makeatother
\author[1,2]{Edwin Ireson\s}
\author[2]{Mikhail Shifman\s}
\author[3,2,4]{Alexei Yung\s}
\affil[1]{{\small
~~School of Physics and Astronomy, University of Minnesota, Minneapolis, 55455, USA
}}
\affil[2]{{\small ~William I. Fine Theoretical Physics Institute, University of Minnesota, Minneapolis, MN, 55455, USA
}
}
\affil[3]{{\small
~ National Research Center ``Kurchatov Institute'', 
Petersburg Nuclear Physics Institute, Gatchina, St. Petersburg
188300, Russia}}
\affil[4]{{\small
~St. Petersburg State University,
 Universitetskaya nab., St. Petersburg 199034, Russia}
}
\title{\begin{flushright}
{\small TPI-MINN-19/12, UMN-TH3821/19 }
\end{flushright}
{\Large{\textbf{
\rule{0mm}{15mm} Composite Non-Abelian Strings with Grassmannian Models on the World Sheet
}}}}
\date{}
\makeatother
\begin{document}
\begin{titlepage}

\maketitle

\begin{abstract}
Most of the non-Abelian string-vortices studied so far are characterized by two-dimensional \cpn models
with various degrees of supersymmetry on their world sheet.  We generalize this construction to ``composite'' non-Abelian strings supporting the Grassmann $\mathcal{G}(L,M)$
models (here $L+M=N$). The generalization is straightforward and provides, among other results, a simple and transparent way for counting the number of vacua in 
${\mathcal N}=(2,2)$ Grassmannian  model.

\end{abstract}
\end{titlepage}


\section{Introduction}
\label{intro}

The 2D \cpnm non-linear sigma model has recently undergone much analysis, in particular
 appearing as world-sheet theories on the simplest non-Abelian string vortices \cite{HT1,ABEKY,SYmon,HT2},
( see \cite{Trev,Jrev,SYrev,Trev2} for reviews) including its heterotic versions \cite{3sv}.
Non-Abelian BPS strings appear in four-dimensional theories with a U$(N)$ gauge group and a certain scalar Higgs 
potential \cite{HT1,ABEKY,SYmon,HT2} ensuring that
U$(N)_{\rm gauge}$ and SU$(N)_{\rm flavor}$ are spontaneously broken down to the diagonal SU$(N)$ in the vacuum.
Unlike the Abrikosov string, they carry orientational moduli due to the fact that, on the the non-Abelian string solution, 
 the above diagonal symmetry is further broken down to 
\begin{equation}
	 \frac{{\rm SU}(N)}{{\rm SU}(N-1)\times {\rm U}(1)}=\mathbb{CP}(N-1)
\end{equation}
leading to the $\mathbb{CP}(N-1)$ model on the world sheet.

In this paper we will generalize the above construction for a non-Abelian multi-string with  $L$ units of  flux, 
introducing a symmetry breaking pattern of the string solution 
\begin{equation}
\frac{{\rm U}(N)}{{\rm U}(L)\times {\rm U}(M)}=\mathcal{G}(L,M)\,,\qquad L+M = N
\label{lm}
\end{equation}
which leads to the Grassmannian model on the world sheet. In terms of gauge linear sigma models this Grassmannian model
can be described as a two dimensional U$(L)$ gauge theory.
Since the original four-dimensional gauge theory is ${\mathcal N}=2$ and the string is $1/2$-BPS saturated, 
the two-dimensional theory on its world sheet is ${\mathcal N}=(2,2)$.

The world-sheet theory for the $L$-multi-string as a U$(L)$ gauge theory was first obtained in \cite{HT1} using a brane
construction. The dimension of its target space, in other words the number of zero modes promoted to two dimensional fields 
of the composite string, was shown to be
\beq
{\rm dim}{\cal M}_{N,L} = 2LN = 2LM + 2L^2,
\label{totdim}
\eeq
where the first term in the final decomposition is the overall number of the orientational moduli, while $2L^2$ describes
$2L$ positions of $L$  components (i.e. the elementary strings with the unit flux) in the plane orthogonal to the string axis and
their ''relative orientations''.

Attempts to reproduce these results in field theory do not lead to a transparent description of the world sheet theory, 
see \cite{ASY} and 
review \cite{Jrev}. 

In this paper we address a simplified problem. We suppress ``relative orientations'' and positions of the component 
strings assuming that axes of all $L$ strings coincide. One last remaining positional zero mode remains, the position of the collective center of the stack of strings. This degree of freedom decouples from the zero mode dynamics of the gauge sector in sufficiently supersymmetric settings: certain deformations of the 4d theory (and consequently the 2d theory also) may couple the positional and internal fermionic zero modes (see \cite{Shifman:2008wv} for a review), it is sufficient to assume $(2,2)$ worldsheet supersymmetry to ensure they will never couple.  The dimension of this reduced  moduli space is 
\beq
{\rm dim}{\cal M}_{N,L}^{\rm reduced} = 2L(N-L)\equiv 2LM
\label{dimreduced}
\eeq 
(see (\ref{lm})) coincides with the dimension of the Grassmannian.
In this setup we  construct explicit multi-string solution and derive  U$(L)$ gauge linear sigma model on
the string world sheet.
We then detail its vacuum structure and check that it coincides with known exact results for such theories.

The paper is organized as follows. In Sec.~\ref{section:42} we introduce four dimensional ${\mathcal N}=2$ SQCD and in 
Sec.~3 construct the solution for a composite string. In Sec.~4 we discuss the full world-sheet theory in
the gauge description and study its classical vacua. Sec.~5 contains our conclusions.

\section{From four to two dimensions. Non-Abelian \\Strings}
\label{section:42}

For what follows we need to briefly review the construction of the ``minimal'' non-Abelian strings with the goal of generalizing it to the ``composite'' strings.

We start off in four dimensional $\mathcal{N}=2$ ${\rm U}(N)$ SQCD, with $N_f=N$ flavors.  The field content reduces to two gauge fields, $A_\mu$ and $A^a_\mu$ (one Abelian and the other not) as well as $N$ flavors of squarks in the fundamental representation of the gauge group, $\Phi^k_A$ where $k$ and $A$ are respectively the color and flavor indices\footnote{We will for now forget about the fermionic matter content, which is present but fully determined by (\ref{elll}) through supersymmetry.},
\begin{equation}
{\mathcal L} =\frac{1}{4g^2_N}\left(F^a_{\mu\nu} \right) +  \frac{1}{4g^2_1}\left(F_{\mu\nu} \right)  + |D\Phi|^2 
+  \frac{g^2_2}{2}\tr\left(\Phi^\dagger T^a \Phi \right) + \frac{g_1^2}{8}\left(\tr\left(\Phi^\dagger \Phi \right) - N\xi  \right) ^2
\label{elll}
\end{equation}
The bosonic action above is a simplified version of the actual bosonic action of  $\mathcal{N}=2$ SQCD. The 
vector supermultiplet also contains scalar complex superpartners of gauge fields, $A_\mu$ and $A^a_\mu$, while squark
fields are described by two sets of scalars, $\Phi^k_A$ and $\tilde{\Phi}_k^A$.  We dropped both adjoint matter  
and squark $\tilde{\Phi}_k^A$ for simplicity as well as associated $F$ terms in \eqref{elll} because these fields 
have no VEVs and play no role in the string solution,
see review \cite{SYrev}. 

The gauge symmetry becomes spontaneously broken (Higgsed) by the introduction of a Fayet-Iliopoulos term $\xi$. 
The scalar equations of motion show that the field $\Phi$ gains a diagonal VEV, enforcing a color-flavor locked phase
\begin{equation}
	\left.\left( \Phi^k_A\right) \right|_{\text{vac}}= \sqrt{\xi} \{\mathbb{1}\}^k_A\,.
	\label{four}
\end{equation}
This means the ground state  is invariant under locked color-flavor transformations ${\rm SU}(N)_{\text{diag}}$.  Topological  line defects, i.e.  
non-Abelian vortices exist
because 
\begin{equation}
\pi_1\left(\frac{{\rm U}(1)\times {\rm SU}(N)}{\mathbb{Z}_{{N}}} \right) =\mathbb{Z}_{{N}}
\end{equation}
where ${\mathbb{Z}}_{N}$ in the denominator stands for the center of SU($N$).
Each of these $N$ solutions are easy to find: one can wind separately any individual element on the diagonal in Eq. (\ref{four})  as we go around a large circle in the perpendicular plane,  introducing one unit of magnetic flux. The U(1) and SU$(N)$ gauge fields are rotated correspondingly. The resulting string has 
tension
\begin{equation}
	T=2\pi  \xi\,,
	\label{simpls}
\end{equation}
to be compared with the tension 
\beq
T_{\rm ANO}=2\pi N \xi
\label{tabr}
\eeq
 of the Abrikosov-Nielsen-Olesen string \cite{ANO} 
in which all of the $N$ flavours contribute magnetic flux to the vortex. 

While this example is simple, it seems generalizable: one may ask what happens if, instead of rotating a single diagonal element in (\ref{four}),
we wind, say, two elements, leaving $N-2$ unwound. In the general case we can wind $L$ elements in (\ref{four}) combining the action of the U(1) generator with the action
of $L$ generators from the Cartan subalgebra of SU($N$),
\begin{equation}
	\left.\left( \Phi^k_A\right) \right|_{\text{large\,circle}}= \sqrt{\xi} \left(\begin{array}{ccccc}
	1&0&0&0&0\\
	0&1&0&0&0\\
	0&0&1&0&0\\
	0&0&0&e^{i\theta}&0\\
	0&0&0&0&e^{i\theta}
	\end{array}
	\right) .
	\label{seven}
\end{equation}
In the example above, Eq. (\ref{seven}), we have $L=2$,  $M=3$, and $N=5$. Moreover, $\theta $ is the polar angle in the orthogonal plane.

Needless to say that in this construction the symmetry of $N$ elements of the ${\rm U}(N)$  Cartan subalgebra under permutations is broken. This
will lead us to a non-Abelian string
with the $\mathcal{G}(L,M)$ Grassmann model on the world sheet. We expect its tension will be 
\begin{equation}
	T_L =2\pi L\xi.
\end{equation}
The symmetry under
\beq 
L\leftrightarrow M
\label{eight}
\eeq
 in the tension is realized in a curious way, namely,
\beq
T_L+T_M = 0\,\, {\rm mod}( T_{\rm ANO})\,.
\label{tlm}
\eeq

 We can remark here that we have explicitly chosen the lower $L$ components to bear magnetic flux, but this choice is arbitrary: any $L$ of the $N$ components can be turned on, there are exactly ${N \choose L}$ such choices, corresponding to different solutions.
The number of distinct strings (\ref{seven}) reduces to the combinatorial coefficient,
\beq
\nu_{L,M} =\left(\begin{array}{c}
N\\
L
\end{array}\right) =\frac{N!}{L! \, M!}\,,
\label{nine}
\eeq
whose symmetry under (\ref{eight}) is evident. We will return to this question later.

\section{Building the ``composite'' string vortex}

The {\em Ansatz} for the string solution in a regular gauge has the form
\begin{align}
\Phi^k_A = U\left(\begin{array}{ccc|ccc}
\Phi_{M}(r)&  &  &  &  &  \\ 
&  \dots&  &  & 0 &  \\ 
&  & \Phi_{M}(r)&  &  &  \\
\hline 
&  &  & e^{i\theta}\Phi_L(r) &  &  \\ 
& 0 &  &  &\dots  &  \\ 
&  &  &  &  & e^{i\theta}\Phi_L(r)
\end{array}  \right) U^\dagger\,,
\label{twelve}
\end{align}
\begin{align}
A^a_{\ell} T^a =\frac{1}{N} U\left(\begin{array}{ccc|ccc}
L&  &  &  &  &  \\ 
&  \dots&  &  & 0 &  \\ 
&  & L&  &  &  \\
\hline 
&  &  & -M&  &  \\ 
& 0 &  &  &\dots  &  \\ 
&  &  &  &  & -M
\end{array}  \right) U^\dagger\,\partial_\ell \theta \left(-1+f_{N}(r) \right),
\end{align}
\begin{equation}
A_{\ell} = \frac{L}{N}\partial_\ell\theta\left(1-f(r) \right) \,, \quad \ell =1,2,
\end{equation}
where $\ell$ denotes spatial coordinates in the perpendicular plane,  $\theta=\arctan\left(x_2/x_1 \right) $ and $r$ is the distance from the string axis in the perpendicular plane. For the time being we ignore the fermion fields: the object we create is BPS protected.
These block-diagonal matrices above are split into a top-left $M\times M$ block and a bottom-right $L\times L$ block, then the non-Abelian gauge potential is indeed traceless with this choice of conventions.

We have introduced four scalar profiles $\Phi_{L},\,\, \Phi_{M}$ and $f_{N},\,\, f$, to be determined later through the equations of motion. Also we have introduced an arbitrary, constant unitary matrix $U\in {\rm SU}(N)$. The scalar functions obey the following boundary conditions required by regularity of the solution at 0:

\begin{align}
&\Phi_{L}(0)=0,\quad f_N(0)=f(0)=1,\\[2mm]
&\Phi_{L}(\infty)=\Phi_{M}(\infty) = \sqrt{\xi},\quad f_N(0)=f(0)=0\,.
\end{align}

In the regular gauge, it is clear  that indeed $L$ colors(-flavors) experience winding, a set of scalars have a topological phase factor that depends on the angular coordinate, and the corresponding gauge field produces magnetic flux. However, it is more convenient for the remainder of the discussion to move to the singular gauge. At the cost of making the gauge fields ill-defined at the origin, we can absorb the phases of the $L$ wound scalar fields and make them functions of the radial distance only, without inducing a winding or topological phase on the remainder $M$ other scalars. Our {\em Ansatz} becomes
\begin{align}
\Phi^k_A = U\left(\begin{array}{ccc|ccc}
\Phi_{M}(r)&  &  &  &  &  \\ 
&  \dots&  &  & 0 &  \\ 
&  & \Phi_{M}(r)&  &  &  \\
\hline 
&  &  & \Phi_L(r) &  &  \\ 
& 0 &  &  &\dots  &  \\ 
&  &  &  &  & \Phi_L(r)
\end{array}  \right) U^\dagger\,,
\end{align}
\begin{align}
A^a_{\ell=1,2} T^a =\frac{1}{N} U\left(\begin{array}{ccc|ccc}
L&  &  &  &  &  \\ 
&  \dots&  &  & 0 &  \\ 
&  & L&  &  &  \\
\hline 
&  &  & -M &  &  \\ 
& 0 &  &  &\dots  &  \\ 
&  &  &  &  & -M
\end{array}  \right) U^\dagger\,\partial_\ell \theta f_{N}(r) \,,
\end{align}
\begin{equation}
A_{\ell} = -\frac{L}{N}\partial_\ell\theta f(r)\,,
\end{equation}
with unchanged boundary conditions. To ease the notation, in what follows we will denote the radial profile for the unwound and  wound scalars as
\beq
\phi (r) \equiv \Phi_M (r)\,,\quad \phi_{\rm w} (r)\equiv \Phi_L (r)\,.
\label{nota}
\eeq

Thanks to the fact that for the purpose of the classical solution our model we limit ourselves to the  bosonic reduction of an $\mathcal{N}=2$ supersymmetric theory, the Lagrangian (\ref{elll}) is at the Bogomoln'yi point and hence half of supersymmetry is preserved on the solution. This allows to write first-order BPS equations of motion for the fields, constraining the profile functions. Namely, 
\begin{align}
	\frac{d\phi(r)}{dr} &= \frac{1}{r} \frac{L}{N} (f-f_{N})\phi(r), \label{BPS1}
	\\[2mm]
\frac{d\phi_{\rm w}(r)}{dr}	&=\frac{1}{N r} \Big(L f(r) - M f_N(r)\Big)\phi_{\rm w} (r), 
 \\[2mm]
	\frac{L}{Nr}\frac{df(r)}{dr} &= \frac{g^2}{4}\Big(L\phi^2(r)+M\phi_{\rm w}^2(r) - N\xi\Big),
	\\[2mm]
	\frac{1}{r}\frac{df_N(r)}{dr}&=\frac{g^2}{2}(\phi_{\rm w}^2 - \phi^2)\,. \label{BPS4}
\end{align}
A string satisfying these equations is BPS protected and can be viewed as a composite of $L$ ``elementary'' strings. Indeed, 
compare their tension
$T_L =2\pi L\xi$  with that of the simplest strings \cite{SYrev} given in Eq. (\ref{simpls}). 

In the above {\em Ansatz}, we introduced an arbitrary unitary matrix $U$, parametrizing an infinite family of solutions
for the composite string. On quantum level there is of course no spontaneous SU$(N)$ symmetry breaking in two dimensions and 
 much in the same way as in CP$(N-1)$ model the moduli space is lifted leaving us with discrete vacua. We will
 recover the ${N \choose L}$ vacuum structure  at the classical level introducing twisted masses (see Sec.~4).

What is important for us now is that not every generic matrix $U$ affects the solution of the type (\ref{seven}). Namely,  any element of the form 
\beq
U = \left(\begin{array}{cc}
{\rm U}_M& 0\\[1mm]
0&{\rm U}_L
\end{array}
\right)
\eeq
(where U$_M$ and U$_L$ are unitary matrices of the dimension $M\times M $ and $L\times L$, respectively)
keeps it intact. 
Thus, the space of distinguishable values the matrix $U$ can effectively take is a group coset, the Grassmannian space
\begin{equation}
	\mathcal{G}_{L,M} = \frac{U(N)}{U(L)\times U(M)}\,.
\end{equation}

Let us present our parametrization explicitly.  We decompose $U\in {\rm SU}(N)$ in two rectangular matrices, arranged columnwise, 
\begin{equation}
	U=\left( \begin{array}{c|c}
	W&X
	\end{array}\right) ,\quad U^\dagger = \left( \begin{array}{c}
	W^\dagger  \\\hline X^\dagger
	\end{array}\right) 
\end{equation}
where $X=X_{Ai}$ ($A=1\dots N,\,\,\, i=1\dots L$) is a $N\times L$ rectangular matrix (a collection of $L$ column vectors of height $N$),
\beq
\{X\} = \left(\begin{array}{ccc}
X_{11}& ...&X_{1L}\\[1mm]
X_{21}&...& X_{2L}\\[1mm]
...&...&...\\[1mm]
X_{N1}&...& X_{NL}
\end{array}
\right),
\eeq
 and   $W=W_{Aj}$ ($A=1\dots N,\,\,\,  j=1\dots M$) is a set of  $M$ column vectors of height $N$,
 \beq
\{W\} = \left(\begin{array}{ccc}
W_{11}& ...&W_{1M}\\[1mm]
W_{21}&...& W_{2M}\\[1mm]
...&...&...\\[1mm]
W_{N1}&...& W_{NM}
\end{array}
\right).
\eeq
 Unitarity of the matrix $U$ imposes
\begin{eqnarray}
	X^\dagger_{iA} X^{\phantom{\dagger}}_{Aj} &=& \mathbb{1}_{ij},\quad W^{\dagger}_{nA} W^{\phantom{\dagger}}_{Am} = \mathbb{1}_{nm},\quad W^{\dagger}_{nA}X^{\phantom{\dagger}}_{Aj} = 0,
	\nonumber\\[2mm] 
	i,j &=& 1,..., L\,,\quad n,m =1, ..., M\,,\,. 
	\label{unitm}
\end{eqnarray}
With these choices the non-Abelian gauge and scalar fields  can then be written
\begin{eqnarray}
A^{{\rm SU}(N)A}_{\ell,B}&=&\partial_\ell \theta\,  f_N(r)\left( \frac{L}{N}\mathbb{1}_{B}^A 
- X^{\phantom{\dagger}Ai} X^\dagger_{iB} \right) \, ,
\nonumber\\[2mm]
\Phi^A_B &=&  \delta^A_B \,\Phi_M(r) + [\Phi_L(r)-\Phi_M(r)] X^{Ai}X^\dagger_{iB}\,.
\label{solution}
\end{eqnarray}
The matrix $W$ drops out of the solution, only $X$ remains as an orientation moduli matrix which should, all constraints and invariances enforced, point in a specific direction in the Grassmannian space. Had we imposed that the upper block of $\Phi^k_A$ in (\ref{twelve}) of size $M\times M $ experienced winding, but not the lower block, then conversely $X$ would drop out and $W$ would become the orientation moduli matrix. Curiously, in either case both $W$ and $X$ will be constrained to live inside the same Grassmannian space.

Let us count the number of the internal degrees of freedom: the Grassmannian space has real dimension
\begin{equation}
{\rm dim } \, \mathcal{G}_{L,M} = 2 LM=2L(N-L)\,,
\label{dim1}
\end{equation}
while $X$'s dimension, as a rectangular matrix with the above constraint, is 
\begin{equation}
{\rm dim } \,  \{ X \} =2LN - L^2=L(2N-L).
\label{dim2}
\end{equation}
These are not the same, however, this is an illusory discrepancy, we have an over-counting of the actual degrees of freedom in $ \{ X \}$. There is an unaccounted-for gauge invariance that we must include. To see this we can turn on dynamics for the $X$ coordinate, and observe gauge invariance of the effective theory that orchestrates its dynamics, but thinking geometrically about the space gives us a hint of why there must be a gauge invariance at hand. 

The Grassmannian space is the space of all $L$-dimensional planes inside $\mathbb{C}^N$ and each point in the space is an individual plane. By choosing a specific $X$ coordinate, we identify this as specifying an orthonormal basis of vectors for such a $L$-dimensional plane. However, this does not describe a unique plane: many such bases, even when constrained to be orthonormal, span the same space. They are all related by change of basis formulae, involving rotation matrices that are elements of U($L$). In field theory terms, this is a continuous symmetry and we expect U($L$) to be a symmetry of the Lagrangian. Now, since it is the physical translation of an over-counting of the degrees of freedom corresponding to physically distinct states, we expect that U($L$) symmetry on the worldsheet is in fact a gauge symmetry. We can prove that this is the case, as mentioned, by explicitly constructing a gauge-invariant action for the $X$ fields.

To realize it we can assume that $X$, previously a constant matrix,  depends on the world-sheet coordinates of the 
string $\tilde{\ell}=(x^0,x^3)$ and generalize the derivation of the world-sheet effective theory for the minimal
non-Abelian string with unit flux,
see \cite{SYrev}. Gauge invariance of the bulk theory is then conserved so long as we turn on some 
extra gauge components, to wit
\beqn
 \left( A^{{\rm SU}(N)}_{\tilde{\ell} =0,3}\right)^{AB} \!\!\!\! &=& -i \left\{ X^{\phantom{\dagger}}_{Ai} \left( \partial_{\tilde{\ell}}  X^\dagger_{iB}\right)  -\left( \partial_{\tilde{\ell}} X^{\phantom{\dagger}}_{Ai}\right)   X^\dagger_{iB} \right. \nonumber\\[1mm]
 &+& \left.X^{\phantom{\dagger}}_{Ai} \left( X^\dagger_{iC} \left( \partial_{\tilde{\ell}} X^{\phantom{\dagger}}_{Cj}\right)  - \left( \partial_{\tilde{\ell}} X^\dagger_{iC} \right) X^{\phantom{\dagger}}_{Cj}  \right) X^\dagger_{jB} \right\} \rho(r) \\[1mm]
 & =&-i \left\{X \left( \partial_{\tilde{\ell}} X^\dagger\right)  -\left( \partial_{\tilde{\ell}} X\right)   X^\dagger+ X\left( X^\dagger\left( \partial_{\tilde{\ell}} X\right)  - \left( \partial_{\tilde{\ell}} X^\dagger \right) X  \right) X^\dagger \right\} \rho(r)\nonumber
\eeqn
for some arbitrary radial profile $\rho$. The latter should obey the following boundary conditions:
\begin{equation}
	\rho(0)=1,\quad \rho(\infty)=0\,,
\end{equation}
the second of which is obvious, though the former will be justified later. Note that the second line makes use of a notational shorthand that will help alleviate the equations we write in the future: by enforcing the rectangular nature of $X$ quite strictly, by always writing $X=X^{\phantom{\dagger}}_{Ai}$ and $X^\dagger = X^\dagger_{iA}$ with the column and row indices ordered this way, the dimensions of multilinear objects composed of $X,X^\dagger$ should never be ambiguous and the matrix products all form intuitively in a neighbor-to-neighbor fashion.

Inserting this full Ansatz in the four-dimensional action and performing the integration over the coordinates transverse to the string axis
produces a two-dimensional  world-sheet effective action for the field $X$, and the addition of the above gauge field not only preserve gauge invariance in the bulk but produces an action which is also gauge invariant on the world sheet, namely,
\begin{equation}
S= \frac{4\pi I}{g_2^2}\int dtdz\,\left( \left|\partial_{\tilde{\ell}} X\right|^2 -\frac{1}{4} \left| X^\dagger \left( \partial_{\tilde{\ell}} X\right)  - \left( \partial_{\tilde{\ell}}X^\dagger \right) X \right|^2 \right) 
\label{wsheet}
\end{equation}
where $X$ is still assumed to be an orthonormal set of vectors (imposed at the level of the functional integration measure, see also Eq. (\ref{action})), and the radial integration constant $I$ is defined by
\beqn
I&=&\int_0^\infty rdr\left( \left( \frac{d\rho}{dr}\right)^2 +\frac{1}{r}f^2_{N}(r)\Big(1-\rho(r)\Big)^2 \right. \nonumber\\[2mm] 
&+& \left. \frac{g^2}{2} \rho(r)^2 (\phi^2_{\rm w} + \phi^2)  -{g^2}\left(1- \rho\right) \left(\phi_{\rm w}  - \phi \right)^2 \right) ,
\eeqn
which is the same expression obtained for the minimal non-Abelian string  \cite{SYrev}.

We see here that $1-\rho$ should vanish as $r$ at the origin in order to cancel the singularity in the term $\frac{1}{r^2}f_{N}$ (in our gauge $f_{N}$ does not vanish at 0). The integral $I$ should be seen as an action for $\rho(r)$ and therefore be varied to determine a minimum of this quantity. This produces an equation of motion for $\rho$ which ties it to the other field profiles. An extremal solution for $\rho$ can be written in closed form in terms of the other profiles
\begin{equation}
	\rho=1-\frac{\phi_{\rm w}}{\phi},\quad \text{for which }I=1\,.
\end{equation}
It is not immediately obvious that this is a solution and requires some algebraic tedium to derive, notably using all of the first-order BPS equations (Eq.(\ref{BPS1})-(\ref{BPS4})) as well a judicious use of integration by parts. It is worth noting that this normalisation constant is \textit{not} $L$, i.e. the total amount of flux running through the string.

The action in Eq. (\ref{wsheet}) has a peculiar and unobvious property. $X$ transforms as a bifundamental 
of U($L$)$\times$ SU($N$), but in fact, through the particular shape of the self-interaction terms, the U($L$) symmetry is made local, despite the absence of any tree-level gauge field. There are several ways of seeing this: one preliminary way of observing this phenomenon also happens to shine light on the $L \longleftrightarrow M$ symmetry that we expect to observe.

Indeed, as we mentioned previously our choice of winding resulted the extra components of the unitary matrix acting on our string solution to vanish: recall that $U$ got split into $W$ and $X$, the latter of which became our basic degree of freedom. We can re-write the Lagrangian we obtain in a way that uses both fields along with a constraint. By using integration by parts we can first rewrite the Lagrangian in Eq.(\ref{wsheet}) as
\begin{equation}
	\left( \partial X^{\phantom{\dagger}}_{Ai} \right) \left( \partial X^\dagger_{iB}\right) \left(\mathbb{1}_{BA} - X^{\phantom{\dagger}}_{Bj} X^\dagger_{jA} \right) 
\end{equation}
Recalling the constraints that unitarity imposes on these matrices, we have that
\begin{equation}
	W^{\phantom{\dagger}}_{Aa} W^\dagger_{aB} + X^{\phantom{\dagger}}_{Ai} X^\dagger_{iB} = \mathbb{1}_{AB},\quad W^\dagger_{aB}X^{\phantom{\dagger}}_{Bi}=\mathbb{0}_{ai}
\end{equation}
alongside orthonormality of $W$, $X$ individually as bases. Assuming we impose all of these constraints in the path integral, we can substitute $W$ back in the Lagrangian above: we obtain
\begin{equation}
		\left( \partial_\mu X^{\phantom{\dagger}}_{Ai} \right) \left( \partial^\mu X^\dagger_{iB}\right)\left(W^{\phantom{\dagger}}_{Ba} W^\dagger_{aA} \right) = \left(W^\dagger_{aA}\partial_\mu  X^{\phantom{\dagger}}_{Ai} \right) \left(\partial^\mu X^\dagger_{iB}W^{\phantom{\dagger}}_{Ba}  \right) 
\end{equation}
It is then a matter of using the mutual orthogonality of $W,X$ to shift derivatives onto the $W$ variables, providing the required re-writing with $W$ as the dynamical variable. 

In addition to shedding light on this issue, this Lagrangian is gauge-invariant under unitary $U(L)$ transformations acting on $X$: perform a gauge transformation
\begin{align}
	&\partial^\mu X^{\phantom{\dagger}}_{Ai} \rightarrow \partial^\mu X^{\phantom{\dagger}}_{Ai} + X^{\phantom{\dagger}}_{Aj} \alpha^\mu_{ji},\nonumber\\
	&W^\dagger_{aA}\partial^\mu  X^{\phantom{\dagger}}_{Ai} \rightarrow  W^\dagger_{aA}\partial^\mu  X^{\phantom{\dagger}}_{Ai} + W^\dagger_{aA}  X^{\phantom{\dagger}}_{Aj} \alpha^\mu_{ji} = W^\dagger_{aA}\partial^\mu  X^{\phantom{\dagger}}_{Ai}.
\end{align}
The local group variation disappears due to mutual orthogonality of $W,X$. Similarly, shifting derivatives onto $W$ we could also discover an $U(M)$ gauge symmetry so long as it is the dynamical variable. Choosing one of the two matrices to have a quadratic kinetic term will hide one of the two gauge invariances. It is surprising that gauge invariance occurs in a theory with no tree-level gauge fields, but we can introduce one to that effect.

We can, in fact, make this accidental gauge symmetry explicit by introducing an auxiliary gauge field $\left( A_{\tilde{\ell}}\right)_{ij}$ in the adjoint of $U(L)$ for the minimal action for $X$ (assuming $X$ is taken to be the fundamental degree of freedom, without loss of generality),
\begin{equation}
	S= \frac{4\pi}{g_2^2}\int dtdz\, \left|\left( \mathbb{1}\partial_{\tilde{\ell}}-i A_{\tilde{\ell}}\right)  X\right|^2
\label{action}
\end{equation}
This gauge field $\left(  A_{\tilde{\ell}} \right)_{ij}$  on the world sheet has no kinetic term (classically). Eliminating it via its equation of motion correctly  produces the effective action in Eq.(\ref{wsheet}) obtained by reduction of the 4D theory. In addition, the $X$ fields are constrained to obey certain orthogonality relations, which so far have been assumed to be enacted in the path integral measure. We can exponentiate this constraint and introduce it to the action as a Lagrange multiplier term: this produces a ``Gauged Linear Sigma Model'' (GLSM) i.e. where the degrees of freedom are allowed to exist in a vector space rather than a more complicated manifold, but whose total degrees of freedom are constrained at tree-level by gauge invariance and auxiliary fields

\begin{equation}
S= \frac{4\pi}{g_2^2}\int dtdz\, \left|\left( \mathbb{1}_{ij}\partial_{\tilde{\ell}} -i \left(  A_{\tilde{\ell}} \right)_{ij}\right)  X_{Aj}\right|^2 + D_{ij}\Big(X^{\dagger A}_iX_{Aj} - \mathbb{1}_{ij}\Big)\,.
\end{equation}
The rotation over the $i$ index in $X_{Ai}$ becomes gauged. Since $X$ is now a linear field, unconstrained in the path integral, it is good to canonically normalise its kinetic term. By rescaling $D$ at the same time this produces

\begin{equation}
S= \int dtdz\, \left|\left( \mathbb{1}_{ij}\partial_{\tilde{\ell}} -i \left(  A_{\tilde{\ell}} \right)_{ij}\right)  X_{Aj}\right|^2 + D_{ij}\Big(X^{\dagger A}_iX_{Aj} - \frac{4\pi}{g_2^2}\mathbb{1}_{ij}\Big)\,.
\end{equation}
This verifies our previous assertion: despite the fact that $X$ now exists in a linear representation of SU($N$)$\times$ U($L$),  U($L$) is in fact not a global invariance of the {\em Ansatz} we presented above but a local one. In addition to the orthonormality constraint, this diminishes the number of real degrees of freedom contained in $X$ by just the right amount: Eq. (\ref{dim2}) effectively reduces to (\ref{dim1}). 
The exact same procedure, in the case when $W$ as the basic degree of freedom, proves that $W$ and $X$ do indeed live in the same space and have the same number of degrees of freedom after gauging the corresponding symmetries: in the case of $W$ we gauge the U($M$) index $m$ so that $2LM +M^2 \to 2LM$.

In this process we may remark that setting $L=1$ produces the minimal non-Abelian string \cite{ABEKY,SYrev}, which has a moduli space based on \cpnm\!\!, a special case of the Grassmannian space. It is in this sense that we call the construction we have outlined a composite string: we can then view the above setup as a synthetic object obtained by fusing $L$  minimal non-Abelian strings (each of string tension $T=2\pi\xi$, the lowest attainable) each with a different color of magnetic flux. Each comes with its own \cpnm internal degrees of freedom, but once the strings fuse and are superposed, these become mutually indistinguishable: this reproduces another possible definition  of the Grassmannian manifold \cite{Ce-Va}
\begin{equation}
\mathcal{G}(L,M) =\left(\mathbb{CP}(N) \right)^L /\mkern-6mu/ S_L
\end{equation}
where $S_L$ is the discrete symmetric group freely interchanging the $L$ copies of 
$\mathbb{CP}(N)$. It is important to emphasize that this is \textit{not} the $S_L$ orbifold of $L$ copies of  $\mathbb{CP}(N)$, since the latter does not have the same dimension as the space that we consider. Rather, it is the set of all maximal orbits under the action of $S_L$, it removes from the space any set of points bearing any definite symmetry included in the symmetric group $S_L$: a point left invariant under any transformation of $S_L$ will necessarily have a shorter orbit and is eliminated from this construction.

From this Lagrangian we can move to a genuine non-linear sigma model, free of constraints and gauge symmetry at the expense of introducing fields evolving in a curved manifold. This involves solving the constraint equation above and produces a metric most analogous to the \cpn Fubini-Study metric. We will briefly detail it here.

We explicitly solve the constraint equation and fix a gauge condition by writing
\begin{equation}
X_{Aj}= \left(\begin{array}{c}
	\varphi_{mi}\\ 
	\mathbb{1}_{ki}
\end{array}  \right) \left( \frac{1}{\sqrt{\mathbb{1} + \varphi^\dagger \varphi}}\right)_{ij}\,,\quad A=1\dots N,\quad m=1\dots M,\quad k=1\dots L,
\end{equation}
introducing $ML $ complex scalars $\varphi=\varphi_{mi}$. These degrees of freedom now directly specify a unique point in the Grassmannian space. Substituting this decomposition into the un-gauged form of the action, Eq.(\ref{wsheet}), we obtain after some algebra very reminiscent of the \cpn model the following Lagrangian, a non-linear sigma model with a metric which generalizes the Fubini-Study metric of \cpn:
\begin{equation}
	\mathcal{L}=\partial_{\tilde{\ell}} \varphi^\dagger_{im}\partial_{\tilde{\ell}} \varphi_{mj}\left( \frac{1}{1+\varphi^\dagger\varphi}\right)_{ji} - \left( \varphi^\dagger_{im}\partial_{\tilde{\ell}} \varphi^{\phantom{\dagger}}_{mj}\right) \left( \partial_{\tilde{\ell}} \varphi^\dagger_{jm}\varphi^{\phantom{\dagger}}_{mk}\right) \left(\frac{1}{1+\varphi^\dagger\varphi} \right)_{ki} \,.
	\label{FSmetric1}
\end{equation}
This can be rewritten in a more symmetric form that treats the $L$-sized and $M$-sized indices equivalently as the following:
\begin{equation}
\mathcal{L}=\left( \frac{1}{\mathbb{1}+\varphi^\dagger\varphi}\right)_{ji}\left( \partial_{\tilde{\ell}} \varphi^\dagger\right) _{im} \left( \frac{1}{\mathbb{1}+\varphi\varphi^\dagger}\right)_{mn}
\left( \partial_{\tilde{\ell}} \varphi\right)_{nj}\,.
 \label{FSmetric2}
\end{equation}

In these forms the target space geometry is made explicit and its properties can be explored in all the usual ways. This manifold is K\"{a}hler, the metric results from the K\"ahler potential
\begin{equation}
K(\varphi,\bar{\varphi}) = \tr \log\left(\mathbb{1}+\varphi \varphi^\dagger \right) \,.
\end{equation}
Given the  K\"ahler potential above, it is straightforward to write a supersymmetric extension for  this non-linear formulation. Our theory, in fact, should be ${\mathcal N}=(2,2)$ supersymmetric, since it is a BPS object. For field theory purposes however we would like to keep the Linear Gauged presentation of the action if possible, which it eminently is.

\section{Supersymmetric Grassmannian model}
\label{sgm}

\subsection{Introducing the full Lagrangian}
\label{itfl}

The action we have derived in the previous section, see Eq.(\ref{action}), is the bosonic, non-supersymmetric version of the Grassmannian model. In practice, many extra fields (including bosonic ones) need to be added in order to get the actual supersymmetric action of the worldsheet theory which should preserve four supercharges.

Let us introduce two superfields, $\Xi^A_i$ and $V_{ij}=V \mathbb{1}_{ij}+V^a T^a_{ij}$, respectively the matter and gauge multiplets, the latter is valued in the Lie Algebra of U($L$). Then, schematically,
\beqn
	\Xi_{Ai} &=& X_{Ai} + \theta \xi_{Ai} + \theta^2 F_{^Ai}\,,\nonumber\\[3mm]
	V&=& \cdots+ \bar{\theta}\theta\left( \sigma^1+i\sigma^2\right) + \theta \sigma^\mu \bar{\theta} A_\mu +\bar{\theta}^2\theta\chi + \bar{\theta}^2\theta^2 D\,,
\eeqn
see Eq. (\ref{sigmas}) for the definition of the $\sigma$ fields.
We combine these in the following superspace action
\begin{equation}
\int d^2x d^2\theta d^2\bar{\theta}\quad \left\lbrace\tr\left( \left( \Xi^A\right)^\dagger e^V  \Xi^A\right)  + \frac{4\pi}{g_2^2}\tr V \right\rbrace  
\end{equation}
The last term is a Fayet-Iliopoulos term. Out of superspace it produces the following Lagrangian:
\beqn
\mathcal{L}
&=&
\left(D_\mu X_{iA} \right)^{\dagger}(D^\mu X_{Ai})  - D_{ij}\left( (X^\dagger_{iA} X^{\phantom{\dagger}}_{Aj}) - 
\frac{4\pi}{g_2^2}\mathbb{1}_{ij}\right) + \bar{\xi}_{iA} (\slashD\xi )_{Ai} 
\nonumber\\
&+&
 \left(\left( i\sqrt{2}\bar{\chi}X\xi\right) +i\sqrt{2}\bar{\xi}^A_i(\sigma^1_{ij}+i\sigma^2_{ij} \gamma^5)\xi^A_j  + \text{h.c.}\right)  -2 (X_{Ai})^\dagger (\bar{\sigma}\sigma)_{ij} X_{Aj}\,.
\label{ws}
\eeqn
Note that $g_2^2$ is the four-dimensional coupling constant. It occurs in our two-dimensional model in the form $4\pi/g_2^2$.

The $D$ coupling is now entirely fixed by supersymmetry, since this auxiliary field is no longer introduced by hand but exists as the top component of the gauge superfield. We can project $D$ into a trace and traceless component, transforming the potential in the following way: define $T^a_{ij}$ to be the generators of SU($L$), then
\begin{equation}
D_{ij}\left( (X^\dagger_{iA} X^{\phantom{\dagger}}_{Aj}) - \frac{4\pi}{g_2^2}\mathbb{1}_{ij}\right) = D\left( X^\dagger X 
- \frac{4\pi L}{g_2^2}\right) + D^a \left( X^{\phantom{\dagger}}_{Aj} T^a_{ji} X^\dagger_{iA} \right) 
\end{equation}

Additionally, an extra bosonic field has been introduced: the gauge multiplet scalar field $\sigma_{ij}$, a matrix of real dimension $L^2$, also in the adjoint representation of ${\rm U}(L)= {\rm U}(1) \times {\rm SU}(L)$, which we expand in real and imaginary components
\beq
\sigma_{ij} =\sigma_{ij}^1 +i \sigma_{ij}^2\,.
\label{sigmas}
\eeq

Note also the structure of the coupling of the gauge scalar to the matter fermions: the appearance of a $\gamma^5$ coupling will yield an anomaly which breaks chiral symmetry.

The auxiliary $F$ terms  in the matter multiplet $\Xi$ vanishes since no superpotential is included. The entirety of the vector superfield is auxiliary -- it has no tree-level kinetic term.

As was explained in Sec. \ref{intro},  in the present paper we address a simplified problem describing the subspace of  
dimension $2LM$ of the full moduli space which has dimension $2LN$, see \eqref{totdim}, \eqref{dimreduced}. In
the Hanany-Tong approach based on the brane picture \cite{HT1} the U($L$) gauge theory on the world sheet contains 
an additional matter multiplet, namely an adjoint multiplet $Z_{ij}$. In particular, the $D$-flatness constraint includes this multiplet and takes the form 
\begin{equation}
D_{ij}\left( (X^\dagger_{iA} X^{\phantom{\dagger}}_{Aj}) +\left[Z^\dagger_{ik},Z^{\phantom{\dagger}}_{kj}\right]
- \frac{4\pi}{g_2^2}\mathbb{1}_{ij}\right).
\end{equation}
The additional adjoint scalar  $Z_{ij}$ describes relative separations of $L$ component strings and their
relative orientations. Our world-sheet theory \eqref{ws} can be obtained from the Hanany-Tong construction in the limit
$Z_{ij}=0$. This limit ensures that all $L$ strings share one and the same axis and they are all ``orthogonal'' 
to each other i.e.
each of the component strings has a flux of a different color.\footnote{ Unlike in other gauge theories that are suggested to arise as low-energy open string modes on stacks of branes, such as $\mathcal{N}=4$ Super Yang-Mills on stacks $D3$ branes, the expectation values of the $\sigma$ field are not related to distances between branes, since it is not a dynamical object, hence the occurrence of an extra field which fits this purpose.}

\begin{figure}[h!]
	\centering
	\includegraphics*[width=0.5\textwidth]{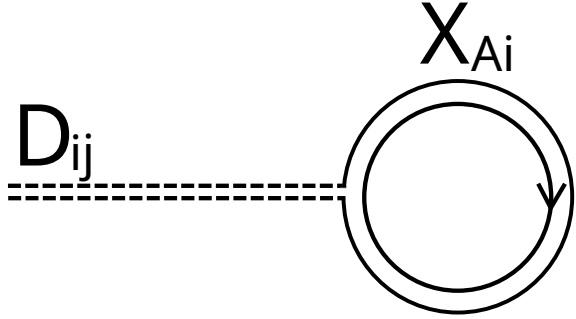}
	\caption{\small The tadpole diagram leading to running of the coupling}
	\label{tadpole}
\end{figure}

The $\beta$-function of Grassmannian spaces, like all other homogeneous coset-type spaces, is known exactly in the supersymmetric case as it is saturated by one-loop diagrams: while it is computed with full precision in the non-linear representation of the theory (particularly since K\"{a}hler manifolds have very regular curvature tensors), in which it appears as a result of target space geometry, a convenient shortcut can be obtained in this representation of the theory by computing the tadpole corrections to an external $D$ insertion, since $\frac{4\pi}{g_2^2}\delta_{ij}$ is the operator coupling to $D_{ij}$ in the Lagrangian. We show in Fig. \ref{tadpole}, drawn in the t'Hooft double-index prescription, how it comes about.

The diagram produces the following loop integral
\begin{equation}
-N  \delta^{ij}\int\frac{d^2q}{(2\pi^2)} \frac{1}{q^2}
\end{equation}
which leads to the following result
\begin{equation}
\beta(g_2^2) = - \frac{N}{4\pi} g_2^4
\end{equation}

This correctly extends the result for $\mathbb{CP}(N-1)$, since it is independent of $L$.  Due to this result, the theory is asymptotically free and is expected to generate a mass scale dynamically, namely, 
\begin{equation}
	\Lambda = M_{\rm UV}e^{- \frac{8\pi^2}{N g_2^2} }\,.
	\label{lambd}
\end{equation}
The dynamical scale parameter $\Lambda$ on the left-hand side is renormalization-group invariant.

\subsection{Vacuum Structure of the Grassmannian String}
\label{cvs}

We can count the corresponding vacua in the classical approximation  by deforming the theory with masses, assumed large compared to the dynamically generated scale. Giving masses to the four-dimensional fields $\Phi$ produces masses for the world-sheet degrees of freedom \cite{SYmon,HT2}: we introduce a set of (complex) masses $m_A$ for each flavor $X_A$ at hand, leading us to us the following bosonic Lagrangian
\beqn
	\mathcal{L}&=&\left(D_\mu X_{Ai} \right)^{\dagger}(D^\mu X_{Ai})
- D_{ij}\left( (X^\dagger_{iA} X^{\phantom{\dagger}}_{Aj}) - \frac{1}{g_2^2}\mathbb{1}_{ij}\right)
\nonumber\\[2mm]
	& -&    
	 2 \sum_{A=1}^{N} (X_{Ai})^\dagger \left( (\bar{\sigma}_{ik}-\bar{m}_A\delta_{ik}\right)\left(\sigma_{kj} 
- m_A \delta_{kj}\right) X_{Aj} \,.
	\label{vaceq}
\eeqn
where 
\beq
\{m_A\}\,, \quad A=1,2, ..., N
\eeq
is a set of $N$ complex twisted mass parameters.
We assume that the masses $m_A$ are all different, $m_A\neq m_B$ for all $A\neq B$, so that
 ${\rm SU}(N)$ is broken  down to U(1)$^{N-1}$. This lifts the orientational moduli
 and allows this theory to isolate its vacua, see review \cite{SYrev} for a similar deformation for the CP$(N-1)$
model on the minimal non-Abelian string. 

The vacua have to satisfy two constraints simultaneously: the orthonormality relations due to the $D$-term potential and the $\sigma$ equations of motion. The two are inextricably linked. 

The former of the two has the following expression
\begin{equation}
	X^\dagger_{iA} X^{\phantom{\dagger}}_{Aj} = \frac{4\pi}{g_2^2} \mathbb{1}_{ij}\,.
\end{equation}
Needless to say, all fermion fields, as well as the kinetic terms, vanish in the vacuum. The vanishing in the vacuum of the last term in (\ref{vaceq})  is the dynamical requirement. 

Inequivalent classical vacuum solutions are described by the expectation values of the fields $X$ and $\sigma$. They are obtained as follows.
From the set of $N$ masses let us choose $L$ of them,
\beq
m_{A_1} ,\, m_{A_2}, \, ..., m_{A_L}\,.
\label{55}
\eeq
These will provide vacuum expectation values for the $\sigma$ field. In the vacuum, only the diagonal elements of the field $\sigma$ are non-vanishing, and they are
\beq
\sigma_{11} = m_{A_1} ,\, \sigma_{22} = m_{A_2},\, ..., \sigma_{LL} = m_{A_L}\,.
\label{56}
\eeq
It is essential that not only the trace component of the  scalar field $\sigma$ gains a VEV, but also  that all the components corresponding to the Cartan subalgebra generators of $SU(L)$ do so as well.
The non-vanishing elements of $X_{Ai}$ must be taken as 
\beq
X_{A_1,1} =\frac{\sqrt{4\pi}}{g_2},\, \, X_{A_2,2} =\frac{\sqrt{4\pi}}{g_2},\, \, ..., 
X_{A_L,L} =\frac{\sqrt{4\pi}}{g_2}\, 
,
\label{57}
\eeq
which corresponds to different choices of winding flavors in \eqref{seven}, see the solution for the squarks
fields \eqref{solution}.

All other components of the matrix fields $\sigma_{ij}$ and $X_{Ai}$ are put to zero. The above solution broadly behaves like $L$ copies of the construction of the isolated vacua of 
\cpn model with twisted masses: this is to be expected, since as we explained previously,
\begin{equation}
\mathcal{G}(L,M) =\left(\mathbb{CP}(N) \right)^L /\mkern-6mu/ S_L.
\end{equation}
In fact, this construction of the Grassmannian space forbids us from taking two different diagonal elements $\sigma_{ii}$ to equal the same mass, a condition without which the counting of the vacua fails to produce the right answer. Indeed, with this criterion the number of the classical vacuum solutions is then obviously
\beq
\nu_{L,M} = {N \choose L} = \frac{N!}{L! \, M!}\,.
\label{59}
\eeq 
given we are choosing $L$ distinct masses for the eigenvalues of $\sigma$.

This is to be compared with the Witten index of the theory \cite{Witten:2000nv}. It is a topological invariant defined by
\begin{equation}
	I_\mathcal{W} = \tr\left(  (-1)^F e^{-\beta H}\right) 
\end{equation}
where we trace over all states in the theory, $H$ is the Hamiltonian derived from the action and $F$ is the fermion number operator (i.e. it weights fermionic states and bosonic states with a sign difference).  It was shown, most generally, that for the K\"ahlerian (Einstein) non-linear sigma models the Witten index is exactly the Euler characteristic of the manifold. By using the explicit target space geometry in Eqs.(\ref{FSmetric1}), (\ref{FSmetric2}) the characteristic can be computed and be shown to match our result quoted in Eq.(\ref{59}). Since this index is a topological invariant we can hypothesize that the number of vacua remains the same in the quantum theory. 

The exact result for the vacuum values of the $\sigma$ fields generalizing the classical expression (\ref{56}) can be inferred e.g. from \cite{ns}. In the full quantum theory, the corresponding equations can be written in the form
\beqn
&&\prod_{A=1}^N \left(\sigma_{jj} - m_A\right) =\Lambda^N \,,\quad \mbox{no summation over}\,\, j, \\[2mm]
&&j =1,2,...,L. \nonumber
\eeqn
As in the classical approximation all off-diagonal values of $\sigma_{jk}$ ($j\neq k$) can be put to zero. The above system of $L$ equations 
can be readily solved in the ${\mathbb Z}_N$-symmetric twisted masses,
\beq
m_k = m_0\exp\left(2\pi \,i \frac{k}{N}\right), \quad k=1,2,...N\,.
\eeq 
In this case 
\beq
\sigma_{jj} = \left| m_0^N+\Lambda^N
\right|^{1/N}\,\exp\left(2\pi \,i \frac{k_j}{N}\right),
\label{68}
\eeq
which matches Eq. (\ref{56}) in the limit $m_0\gg\Lambda$. When $m_0$ is set to zero, this formula also provides the mass spectrum for the low-lying excitations of the theory. Indeed, the entire action can be put into Landau-Ginsburg form \cite{Ce-Va}, in which Eq.(\ref{68}) appears as the equation that minimizes the superpotential appearing in this formulation. Thereby, the low-lying energy states are kinks interpolating between minima of this potential. We label each minimum by its set of masses
\begin{equation}
	V_{\lbrace k_1,\dots k_L\rbrace}=\ket{\sigma_{11}=\Lambda e^{2\pi \,i \frac{k_1}{N}},\dots\sigma_{LL}= \Lambda e^{2\pi \,i \frac{k_L}{N}}}.
\end{equation}
This vacuum is independent of the ordering of the $k_i$, it is a function of the \textit{set of values} rather than the values themselves.

Fundamental solitons will exist between two vacua whose set of indices differ in only one element: $V_{\lbrace k_1,\dots k_L\rbrace},\,	V_{\lbrace k'_1,\dots k'_L\rbrace}$ will connect if\footnote{Without loss of generality, since ordering of the indices does not matter, we take the unequal elements to have the same index by relabeling and reordering them.}
\begin{equation}
	k_1 = k'_1, \,\dots \quad k_j - k'_j = r \neq 0 ,\, \dots \quad k_L = k'_L.
\end{equation}
The index difference $r$ is defined modulo $N$, and the sign of $r$ is irrelevant in what will follow, or, to put it another way, there is a $r\leftrightarrow N-r$ symmetry in the structure of vacua.

The mass of the object interpolating between these two vacua is therefore
\begin{equation}
	m_r = \frac{N}{2\pi}\Lambda \left|\exp\left(\frac{2\pi i k_j}{N}\right)-\exp\left(\frac{2\pi i k'_j}{N}\right)\right|=\frac{1}{\pi} N \Lambda \sin\left( \frac{\pi r}{N}\right) .\label{massf}
\end{equation}
This formula is indeed invariant under the advertised symmetry.

The objects of truly minimal mass are therefore the ones interpolating between vacua where only one pair of indices are unequal and differ only by 1 (mod $N$), which we may call the ``closest neighbors''. Pairs of vacua differing in one index by more than one unit will have higher mass as a result, and vacua ``further away'' with multiple unequal indices are considered not to be fundamentally connected at all \cite{Bourdeau:1994je}. This is different from \cpn, where the latter case does not exist, all vacua are connected to each other in the sense of the above.

\begin{figure}[h!]
	\centering
	\includegraphics[width=0.3\textwidth]{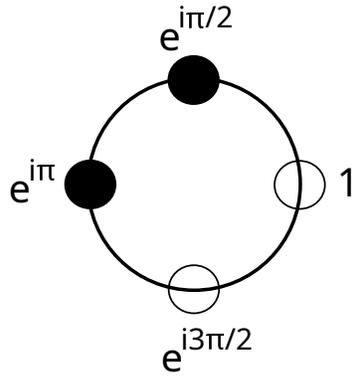}
	\caption{A pictorial representation of a particular vacuum in $\mathcal{G}(2,4)$, one where the eigenvalues of $\sigma$ are given by the roots of unity marked with black dots. The vacuum pictured here is $V_{\lbrace 1,2 \rbrace}$.}
	\label{vacuumexample}
\end{figure}

This notion of connectedness is considered to be an exact one, as this whole discussion can be derived directly from considerations of  topological-antitopological fusion \cite{Ce-Va}. Using these methods, further information can be extracted, for instance the degeneracy of the solitons interpolating between two given vacua. In $\mathbb{CP}(N)$ the lightest solitons in the theory have multiplicity $N$ \cite{HoVa}, all of which having the same mass. Using the topological construction of the vacua the multiplicity of the solitons with mass $m_r$ between two specific vacua was shown to have multiplicity
\begin{equation}
	n_r = {N \choose r},
\end{equation}
which, again, does have the advertised symmetry.

To summarise, the low-lying excitations around these vacua consist, much like in $\mathbb{CP}(N)$, of an $N$-plet of kinks interpolating ``nearest neighbouring'' vacua (i.e. $r=1$), all of which have mass given by the mass formula above in Eq.\ref{massf}.

\begin{figure}[h!]
	\centering
	\includegraphics[width=0.5\textwidth]{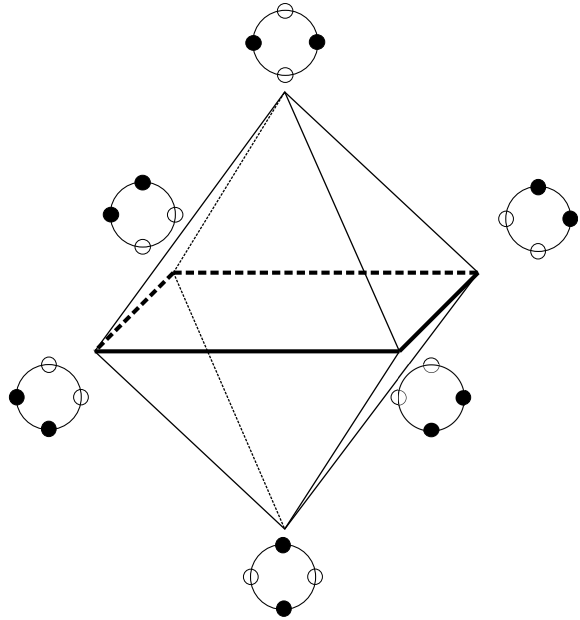}
	\caption{The adjacency graph of the six vacua in the theory. Each vertex of the octahedron is associated to a vacuum state represented by a roots-of-unity diagram. The central square connects neighbors whose indices differ by more than one unit, thus have a higher mass, represented by a thicker line. No soliton, elementary or bound, exists to connect the top and bottom vertices.}
	\label{vacuumgraph}
\end{figure}
\clearpage
Let us present a graphical illustration of the structure of the theory in a simple example. Let us work in the smallest (non-trivial) Grassmannian space
\begin{equation}
	\mathcal{G}(2,4)=\frac{U(4)}{U(2)\times U(2)}
\end{equation}
 There are four possible values that the $\sigma$ eigenvalues can take, and there are two of the latter, so there should be six vacua. Since these values are roots of unity, we can represent an individual vacuum using a unit circle on the complex plane and marking which roots of unity are used up by the $\sigma$ fields, see Fig.\ref{vacuumexample}. Then, we can draw a graph which connects neighboring vacua. For the simple example at hand, we obtain an octahedral structure as shown in Fig.\ref{vacuumgraph}.

Even should we focus exclusively on objects of minimal mass, for which the neighboring vacua differ by one unit in one index, the connectivity of the vacua shows some non-trivial structure, and the polytopes required to display even these ``closest neighbors'' will become complicated very quickly.

\section{Conclusions}

In this work we have constructed the composite non-Abelian vortex string solutions   
and investigated some of their properties. In particular, we derive the world-sheet effective theory
for the reduced number of moduli living on the string. These moduli describe overall orientations
of the composite string inside SU$(N)$ group. Much like its more elementary counterpart, with $\mathbb{CP}(N)$ on its world sheet, this string is topological in nature, is BPS protected, and possesses some leftover gauge degrees of freedom along its worldsheet. These fields live in a generalization of the \cpn space usually seen in elementary non-Abelian vortices, the Grassmannian space, which nonetheless formally looks very similar to \cpn. The vacua of this theory were exhaustively justified through several different means and some aspects of the quantum behaviour were touched upon. 

\section*{Acknowledgments}

The authors are grateful to P. Koroteev for a useful discussion. The work of MS was supported in part by the DOE grant \mbox{DE-SC0011842}. 
The work of A.Y. was  supported by William I. Fine Theoretical Physics Institute,   
University of Minnesota, and 
by Russian Foundation for Basic Research Grant No. 18-02-00048a.

\end{document}